    \documentstyle[11pt,paspconf,epsf]{article}

    \def\etal{{\it et al. }}

\begin{document}
    \title{ Dwarf Irregular Galaxies and the Intergalactic Medium }

        \author{Noah Brosch }

    \affil{The Wise Observatory and
    the School of Physics and Astronomy \\ Tel Aviv University, Tel Aviv 69978,
    Israel}

        \begin{abstract}

    Dwarf galaxies (DGs) are more numerous than large galaxies. 
    Most dwarfs in clusters are dEs, but in the field they belong mostly to late types.
    The importance of late-type DGs in the context of the intergalactic medium (IGM) lies
    in the fact that (at least some) are probably ``young'' galaxies forming
    stars for the first time. Many DGs have significant amounts of
    interstellar gas, they may form out of intergalactic gas clouds, and
    matter ejected from such galaxies as a result of star formation
    processes may enrich the IGM with metals. The physical mechanism responsible
for triggering the star formation process has not yet been identified.
The extended halos of DGs may
    provide (at least some of) the QSO absorption line systems. Recent
    observations also show that the LSB dwarf galaxies may be good tests of
    MOND. 

\end{abstract}

\vspace{3mm}


        \section{Introduction: the panoply of galaxies}

    Taxonomy of phenomena is the first, and most basic, of scientific
    investigations aimed at understanding nature. Astronomers classified
    galaxies since the 1920s, from their images on blue-sensitive
    photographic plates, as belonging to two kinds, ellipticals and spirals.
    Hubble (1936) refined later
    the classification to include barred as well as ``normal'' spirals, and
    introduced different subclasses of E and S galaxies. This
    classification, depicted in the well-known fork diagram, is in use even
    today, albeit with small modifications and with significant additions.

    Hubble classified further the elliptical galaxiess according to their degre of
    flattening. A galaxy belongs to sub-type En if its apparent semi-axes a
    and b form the index n=10(a-b)/a. No elliptical galaxies flatter than
    n=7 exist;  their semi-major axes would be longer by more than 
70\% than the  semi-minor axes  and such objects would be classified as edge-on
    spirals. A spiral is a disk galaxy which can even be fully planar, with
    no bulge at all, and may belong to sub-types a, b, or c, according to
    the openness of its (spiral) arms and the relative strength of the
    central bulge relative to the disk. A galaxy could belong to any of
    these sub-classes and be either barred [SB] or regular [S]. Finally, 
    another type
    of disk galaxies can be as flat as the spirals but may lack spiral arms; such
    galaxies belong to the lenticular category (S0).

    Even in Hubble's original scheme there were outstanding objects which
    did not fit any of these categories. Such examples, visible
    with the naked eye in the southern hemisphere, are the Large and Small
    Magellanic Clouds (LMC \& SMC). These satellites of the Milky Way galaxy
    are irregular galaxies. Both Magellanic Clouds belong to the
    Irr I type and are perhaps failed or tidally-stripped spirals 
    (it is easier to visualize this for the
    LMC than for the SMC). The Irr II class contains the genuinely irregular
    examples, but note that the classification of Irr objects has been
    revised many times.

    A recent classification, in particular of the irregular
    galaxies, is the result of a seminal work by Binggeli, Sandage, and
    Tammann and consists of the most comprehensive morphological
    classification of galaxies in the direction of the Virgo cluster (VC).
    The Virgo cluster galaxy catalog (VCC: Binggeli \etal  1985) contains
    more than 2,000 objects; most are dwarf ellipticals. These objects have
    small dimensions but exhibit the general photometric behavior of
    elliptical galaxies (smooth light distribution, r$^{1/4}$ surface
    brightness profiles, red colors, lack of interstellar matter [ISM]). 
    Almost 300 objects in the
    VCC are late-type dwarf galaxies and all have been observed by Hoffman
    (1987, 1989) with the Arecibo radiotelescope in the 21 cm line. 
    The classification
    scheme introduced by Sandage \etal (1985) for those objects distingushed
    among five sub-classes of dwarf irregular galaxies (DIGs)
    separated by luminosity: the most luminous were
    classified as Irr III and the least luminous as Irr V.

    In addition to the dEs and Irr III-V an additional class of DIGs was
    identified: the blue compact dwarfs (BCDs: see {\it e.g.,} Loose and Thuan
    1986 for a taxonomic study of BCDs). This class is characterized by
    relatively bright objects of very small dimensions, sometimes
    undistinguishable from stellar images except by their spectra. A
    sub-class of BCDs, called ``HII galaxies'' ({\it e.g.,} Telles \etal 1997), is
    characterized by strong emission-line spectra. There is considerable cross-over
    between the BCD and HII galaxy classes, as many BCDs also show strong
    emission lines, but it is clear that in both cases the galaxies have
    considerably high surface brightness. In particular, the ``type II''
    less luminous
    HII galaxies are, in fact, BCDs with a central (nuclear) starburst.
    Also, there are BCDs devoid of emission lines (Almoznino \& Brosch 1998).

    de Vaucouleurs (1994) revised the historical Hubble sequence as a
    three-dimensional distribution. An object could be located in the
    classification volume along three axes: family, variety,
    and stage. The  ``family'' variable distinguishes among ordinary [A],
    barred [B], and transition [AB] types. The second variable can be either
    spiral-shaped [s], ring-shaped [r], or a transition between the two
    [rs]. The third is essentially the T-type used, {\it e.g.,} in RC3; 
    ellipticals and spheroidal galaxies have T=--6 to --4,  S0's have
    T=--3 to --1, and  spirals span the range T=0 (S0/a) to T=9
    (Magellanic Irregulars). The compact late-type galaxies are relegated to
    T=10 and 11.

    The different morphological classifications are linked to physical
    characteristics. From ellipticals to late-type spirals (Sc) the spin of
    a galaxy and its ISM content increase while the total symmetry
    decreases. The irregulars stand out by showing a large variety of gas
    contents. The difference between normal and dwarf galaxies has been set,
    rather arbitrarily, at M$_B$=--18; dwarfs are fainter than this limit.
    The dynamical and ISM properties of galaxies translate into observables
    related to star formation (SF). The more gas-rich a galaxy is the higher
    its star formation rate (SFR) can be, but not all gas-rich galaxies 
form stars, at least not at    present.

    A recently recognized class of galaxies consists of low surface
    brightness (LSB) objects, with central surface brightness below 23.0
    mag./square arcsec. The class has been identified by Bothun and has been
    studied by him and collaborators (see {\it e.g.,} Bothun \etal 1997 for a
    review). The best-known member of this class is the giant LSB disk Malin
    1, but there are many dwarf galaxies which are LSBs. In fact, the dS
    class and all Irr IV and V belong to the LSB kind. The LSBs may
    prove to be an important component of the galaxy population (McGaugh et
    al 1995), and those objects detected so far may be only the ``tip of the
    iceberg'' in their numbers (de Blok 1997). However, Briggs (1997) argued
    that LSBs may contribute only a small fraction of the total HI content
    of the Universe. 

    It is interesting to note that some of the blue galaxies identified in
    the Hubble Deep Field (HDF) are physically small and have high surface
    brightness; they would qualify as extremely high
    surface brightness BCDs, i.e., those with very high SFRs ({\it e.g.,} 
Ferguson \& Babul 1998). Some compact and faint galaxies, seen with the HST at z$\sim$2.4,
    may be sub-galactic entities which will eventually coalesce into 
    a large galaxy (Pascarelle \etal. 1996). These messages from the past 
history of the Universe emphasize that (a) some part of the galaxy
evolution may be  similar to what we now witness in
local DIGs, and (b) distant DIGs are presumably galactic building
blocks of present-day large galaxies.

        \section{Star formation in DIGs and in other galaxies}

    The star formation process in large and small galaxies modifies their
    metal and gas contents. The more massive stars enrich the ISM of a
    galaxy with heavy elements produced by nucleosynthesis and ejected from
    a star by radiation pressure during the red (super)giant stage or during
    a type II supernova (SN) explosion. Additional ISM enrichment, with
    other types of metals, happens when a white dwarf explodes as a type I
    SN or during the evolution of close binaries. This metal-enriched ISM
may provide an important source of metals for the IGM.

    A significant difference in the mode of star formation between DGs and
    large galaxies (LGs) lies in the number of possible triggers of the SF
    process. Whereas the large galaxy SF may be triggered by whole-disk
    processes ({\it e.g.,} shear forces, spiral density waves), these are probably
    absent in DGs. Thus, understanding SF should be simpler in DGs than in
    LGs.

    We studied SF in DGs of the Virgo cluster and the results will be
    discussed here by Almoznino. Briefly, we found that SF takes place in
    bursts and is present in both high surface brightness (HSB) objects as
    well as in low surface brightness ones. Other results from this study
    show that the star-forming regions in DIGs are most often observed at
    the visible edge of a galaxy, particularly to one side of the galaxy.
    This result shall be mentioned again, towards the closing of this
    presentation, in the context of DM and its influence on the star
    formation.

    Matter may be ejected from DGs by the SF activity. This was predicted by
    Dekel \& Silk (1986) and was observed in a number of objects ({\it e.g.,}
    Papaderos \etal 1994, Marlowe \etal 1995; Martin 1997). In some cases,
    the starburst activity heats up the ISM gas to very high temperatures
    and the gas radiates in the X-ray ({\it e.g.,} Della Ceca \etal 1996, 1997 
for ASCA observations; Bomans \& Grant 1998 for ROSAT observation). The
    matter ejection is also observed as major disturbances of the ISM
    dynamics and as extended HI and ionized gas ``ansae'' of a galaxy.
    These may form, in some cases, structures that reach out several kpc
    from the galaxy ({\it e.g.,} Hunter \& Gallagher 1997; Brosch \etal 1998a). 
Note that while the
    inflation of gas bubbles in a galaxy is an established observational
    fact, the question of whether this matter is completely lost by 
the galaxy or whether it is
    retained and eventually recycled because of the influence of an
    extended DM halo has not yet been solved. Recent observations (Meurer
    1998) indicate that it is very difficult to completely clean a galaxy of
    its ISM. NGC 1705, which shows presently a very strong SF burst, may
    eject eventually $\sim$8\% of its ISM during the present event. With the
    accepted bursting rate of one per Gyr, the full gas content may be
    released from the galaxy only over one Hubble time.

    The metal enrichment of the Universe can be traced from a computation of
    its star formation history (Madau 1997). Until very recently it was
    believed that the star formation reached its peak at z=1-2 and that both
    before that time and since z=1 to the present the SF was low. New
    observations with the SCUBA (Submillimetre Common-User Bolometer Array) 
    instrument on the
    JCMT in Hawaii reveal objects with
    sub-mm emission which could be high redshift (z=1.5-3) galaxies. The
    emission is indicative of stellar radiation reprocessed by dust in
    these galaxies. The luminosity corresponds to SFR$\simeq$140-910
    h$^{-2}_{75}$ M$_{\odot}$ of gas per year converted into stars (Berger \etal
    1998); these are
    extremely high SFRs, $\sim$one order of magnitude higher than previous
    maximal estimates. Other IR-bright very distant galaxies were apparently
    identified in deep NICMOS images of the Hubble Deep Field.

    The existence of distant, IR-bright galaxies implies a revision of the 
    Madau (1997) scenario for the
    history of SF in the Universe; whereas Madau showed that the SF peaks at
    z=1-2, the SCUBA observations indicate that (a) the SFR stays $\sim$flat at
    least up to z=3 and perhaps up to z=5, and (b) many high-redshift galaxies
    may be hidden from optical detection by dust extinction. In addition,
    the new results show also how fast  the metal enrichment of a galaxy
    (understood as the production of dust grains) is
    proceeding, because when observing galaxies at z=3-5 one is probing
    look-back times which are only about 4-5 Gyr (H$_0$=50, q$_0$=0), or
    1-2 Gyr (H$_0$=50, q$_0$=0.5), after the Big Bang.

    In this context we may wonder what role can DIGs play in the metal
    enrichment history of the Universe. Although it is not clear whether
    matter escapes at all from a DIG, the matter ejection from DIGs should
    be easier than from larger galaxies. In addition, the reduced (local)
    gravitational potential ensures that any material levitated into the
    halo will remain there for a long time indeed, before sinking back into
    the galactic reservoir of ISM. 
In other words, ``galactic fountains'' originating from DIGs should sprout
higher, and the material fallback should be longer, than in large galaxies.
As the binding energy at high halo
    locations is much lower than in the visible galaxy, the metal-enriched
    material levitated there may be separated more easily from its parent
    galaxy than matter in the optical disk, by {\it e.g.,} ram pressure effects.
    In any case, either proto-galactic bodies may exist and the present-day DIGs
    may form out of these, or matter might
    escape/be torn off dwarf galaxies, resulting in {\bf some} intergalactic
    objects (almost) devoid of stars but containing some baryonic
    non-luminous matter. In both cases, DIGs and their precursors prove
    to be important IGM components.

        \section{Intergalactic absorbers and their connection to DGs}

    The discovery of absorption lines in spectra of high-redshift objects
    (Bahcall 1968), and in particular the existence of lines at a redshift
    very different from that of the object whose spectrum was analyzed,
    emphasized the presence of gas in the intervening space up to the
    highest redshift QSOs. Many of these systems, perhaps most of them, 
may be produced in DIGs and/or in their immediate precursors.

The absorption-line systems have been classified
    into three categories: the Lyman-$\alpha$ systems which show only
    absorption from this line, the Lyman-limit or damped Lyman-$\alpha$
    systems (the Ly-$\alpha$ absorption line in these systems is saturated
    at the line center), and the metal-line systems where additional
    absorption lines, produced by neutral or ionized elements heavier than H
    and He, are observed ({\it e.g.,} Wolfe \etal 1986). All the metal-line
    systems belong also to the damped Ly-$\alpha$ class. The main difference
    between these systems lies in the hydrogen column density [log N(HI)]: this is
    12-15 cm$^{-2}$ for the Ly-$\alpha$ systems, 16-19 cm$^{-2}$ for the
    damped Ly-$\alpha$ systems, and $\geq$20 cm$^{-2}$ for the metal line
    systems, which can reach up to 22-23 cm$^{-2}$.

    The absorption line systems have diverse clustering properties, but
    mostly they tend to follow the large-scale distribution of galaxies. A
    claim that the Ly-$\alpha$ absorbers ``avoid the voids'' has not been
    confirmed by new HST observations (Shull \etal 1996). Ly-$\alpha$ systems  with 
log N(HI)$>$12.7 cm$^{-2}$ are found every $\sim$3,400 km/s. If the entities producing the
    Ly-$\alpha$ absorbtion are spherical bodies with 100 kpc radii, their masses
    are 10$^9$ M$_{\odot}$ and their space density would be comparable to that of
    the dwarf galaxies. This is not the case for the damped Ly-$\alpha$ systems;
    the distribution of their metal abundances with redshift can be reproduced
    with evolutionary models of large spiral galaxies (Ferrini \etal 1997).

    In addition to matter which reveals itself through the optical-UV (rest
    frame) absorption lines one can also hope to detect neutral matter
    through its HI 21 cm line. Many such attempts were made as ``blind
    searches'' for HI signatures ({\it e.g.,} Brosch \& Krumm 1984 with the Green
    Bank 300' radiotelescope [search for emission features], Brosch 1989
    with the Arecibo radiotelescope [search for 21 cm absorption features
    against radio QSOs]), but were unsuccessful. A positive result was
    obtained by Szomoru \etal (1996) in a VLA survey for HI emission from
    the Bootes void, where 18 HI objects not
    previously known were found.

    Apart from (mostly negative) attempts to detect isolated HI emission,
    searches were conducted for HI features associated with galaxies. 
These were successful in detecting a
    giant HI ring around the small group of galaxies in Leo centered on M96
    (Schneider \etal 1983), extended HI emission in the M81 group (Lo \&
    Sargent 1979), HI companions to dwarf galaxies (for $\sim$25\% of the
    cases: Taylor \etal 1996), a large neutral hydrogen clound in the
    southern outskirts of the Virgo cluster (HI 1225+01: Giovanelli \&
    Haynes 1989), and the Malin-1 galaxy.

    The case of Malin-1 is particularly interesting. The object was
    detected originally as an LSB through photographic amplification
    (``Malinization'') of optical images taken by a Schmidt telescope in the
    direction of the Virgo cluster (Bothun \etal 1987). The location of the
    object, which appeared as a faint ``smudge'' on the original plate,
     showed a large galaxy after the photographic amplification. The nuclear
    region showed a faint [OII] emission line at z=0.083, indicating that
    the galaxy was in the far background of the VC. The redshift was
    confirmed later by a 21 cm observation, which showed that the giant LSB
    galaxy is very hydrogen-rich.  Malin-1 is so large that if it would   
     be located  in the VC its amplified optical angular size would  be $\sim$one
    degree !

    Additional searches for LSB galaxies were done by Schombert (1997) on
    the PSS II plates, by photographic amplification of plates of the nearby
    Fornax cluster (Morshidi \etal 1997), by automatic scanning of plates (APM survey,
    Loveday 1997), and by a wide-field CCD survey (O'Neil
    \etal 1997). Until now, only three giant LSB galaxies have been
    identified: Malin 1 (Bothun \etal 1987), F568-6 (Bothun \etal 1990), and
    1226+0105 (Sprayberry \etal 1993). The latter is outstanding in that it
    is located very close (in projected sky position) to the HI 1225+01
    feature mentioned above. Note though that its redshift is 23,660 km
    s$^{-1}$, well in the background of the VC. These searches also
    yielded many tens of ``normal''-sized  LSBs

    It is useful, at this point, to summarize the observed properties of
    LSBs. These galaxies lack bulges, bars, and nuclear activity, as well as
    CO or IR emission (i.e., no molecules or dust). Their typical SFRs are
    $\sim$0.1 M$_{\odot}$/yr and the metallicities are  $\sim$1/3 solar. Their colors are
    generally blue, though some red LSBs have been detected (O'Neil \etal
    1997). The HI rotation curves, measured by de Block \etal (1997) and by
    Pickering \etal (1997), indicate that their gaseous component is
    dynamically significant at all radii and that the galaxies are dark-matter
    (DM)-dominated at all radii; the baryonic component is less than 4\% of
   the total mass. Mass
    models for the LSBs indicate that their DM halos are less dense and more
    extended than DM halos of HSBs. One may extrapolate from this to the LSB
    dwarfs in the VC studied by Heller \etal (1998), in particular so as to
    explain the asymmetry of SF (Brosch \etal 1998b). This could be produced if the galaxy
    orbited on an off-center track within the DM halo with a retrograde spin
    in respect to the sense of its orbital motion (Levine \& Sparke 1998),
    or if the halo itself had some intrinsic anisotropy (Jog 1997).

    The question one can ask at this point is whether there is a clear
    connection between Ly-$\alpha$ absorption systems, HI emission/absorption
    features,
    and dwarf or LSB galaxies. The best test case is offered by Ly-$\alpha$
    absorption lines produced by nearby material and discovered in the UV
    spectrum of 3C273. This high luminosity QSO is at a redshift of 0.158
    and its HST UV spectrum (Bahcall \etal 1991) shows two Ly-$\alpha$
    absorptions produced by material at the redshift of the VC. This is clearer from the
    high-resolution UV spectroscopy (Weymann \etal 1995), where the two
    lines appear at 1,012.4$\pm$2 km/s and 1,582.0$\pm$2 km/s. Both 
Ly-$\alpha$ lines
    have approximately the same column density: log(HI)=14.19$\pm$0.04 and
    14.22$\pm$0.07 respectively. The hydrogen could be associated with the
    large isolated HI cloud mentioned above (HI 1225+01), at 1,298$\pm$20 km
    s$^{-1}$ and which has two concentrations, where the highest HI column density 
peak harbors a dwarf irregular galaxy (Salzer \etal 1991), but its dynamics may
make this explanation problematic.
 
    While it is clear that the Ly-$\alpha$ absorptions are produced by
    hydrogen atoms in the VC, very deep searches for luminous material close
    to the line of sight to 3C273 and which may be associated with this
    intergalactic hydrogen body have been so far unsuccessful. One of the more
    promising explanations (Hoffman \etal  1998) associated one of the
    Ly-$\alpha$ lines with a dwarf irregular galaxy (MCG+00-32-16) rather
    distant from the line of sight to 3C273 ($\sim$200 kpc, for a VC
    distance of 18 Mpc); if this is a typical DIG, then the HI envelopes of
    such galaxies may extend to a large fraction of a Mpc. One should then
    wonder about the mere existence and long-term stability of such 
extended halos in the  gravitational  environment of the VC.

    It is not clear what role, if any, do the HI companions play in the
    development of a DIG. In a number of cases, notably the two galaxies
    with the lowest known metal abundance I Zw 18 (Dufour \etal 1996) and
    SBS0335-052 (Melnick \etal 1992), two optical entities are distinguished
    which may be two separate and interacting galaxies, or may be
    understood as two parts of the same system. In the case of I Zw 18 the
    different stellar concentrations of the galaxy have somewhat different 
    stellar histories (Hunter \& Thronson 1995). In SBS0335-052 two small
    galaxies share the same surrounding HI cloud (Pustilnik \etal 1997). 
    It is interesting that the
    major axis of this cloud points in the same direction as the major axes
    of the two dwarf galaxies, and all three axes point back to a large
    spiral, which is $\sim$150 kpc away and has a similar redshift. A
    similar double structure is seen in HI1225+01, where one lobe harbors a
    star-forming DIG. The three objects, although not forming a statistically significant
    sample, indicate that some sort of interaction may be triggering the SF process in DIGs.

        \section{Puzzling questions}

The consideration of DIGs in the context of their contribution to the
 IGM leaves a number of puzzling questions.
We shall not discuss all, but will single out three, related to the nature of the DM,
the mechanism triggering the SF, and their connection with an alternate theory of gravity.

    Searches for HI not associated with luminous material have, so far, been
    rather unsuccessful, except in the neighborhoods of luminous galaxies. On
    the other hand, there are indications that the Universe contains more
    matter than it is actually observed; this is the dark matter which
    provides part of the binding force of galaxies and clusters of galaxies.
    Until now, no fully suitable explanation of the nature of dark matter has been
    proposed. As mentioned above, LSBs, and in particular dwarf LSBs, may
be DM-dominated at all radii. They may, therefore, be the best case studies for
establishing the nature of the DM.
    One question which may be posed from a study of galaxies, in particular
    those of low surface brightness, is whether there are dark matter halos
    without (neutral) gas and devoid of stars. Such halos may provide seeds
    for the formation of future DGs. This question is linked to and may be
    constrained by lensing arguments; such DM halos would make ideal
    gravitational lenses, being $\sim$massive and $\sim$transparent.

O'Neil \etal (1998) proposed that SF is
    triggered by distant (soft) tidal encounters. This is similar to the mechanism of Icke
(1985), and moreover the
    scenario proposed by O'Neil \etal (1998) seems to produce mainly spiral
    arms and inner ring density enhancements, not one-sided SF at the edge
    of a galaxy as shown in Brosch \etal (1998b). The morphology of SF can,
therefore,  constrain the evolution of galaxies.
  It is also possible to test the hypothesis of tidal triggers of SF with galaxies
from Zaritsky \etal (1997). Their sample contains 115 satellite galaxies in 69
systems, where the primaries are large isolated spiral galaxies. Zaritsky \etal
list for each satellite the (projected) distance from its primary and the
presence or absence of emission lines in its spectrum. The presence of
emission lines  implies a recent SF event (because of the need for ionizing photons
to produce these lines); we check this against the distance of the satellite
from its primary. The expectation is that the closer a satellite is to its
primary galaxy, the stronger the tidal force must be and the stronger should be its SF.

One should be careful in such comparisons because (a) strong and recurrent SF may deplete
a nearby satellite of its ISM preventing further SF, although tides could
still be strong, (b) the listed distances are only projections and an object
which seems nearby may actually be more distant, and (c) there are significant
numbers of satellites with no spectral indications. With all these caveats, we
remark that the sample of Zaritsky \etal does not seem to show a link of the
distance of a satellite from its primary. In particular, the fraction of satellites
  with emission lines among those within 200 kpc of their primaries (85\%) 
is exactly the same as for the
satellites more distant than 200 kpc. I conclude that 
distant tidal interactions are not likely triggers of SF in DIGs.

    Finally, LSB galaxies are apparently an excellent laboratory for testing
    gravity. An interesting proposal by Milgrom (1983) to explain the flat
    rotation curves of disks without invoking DM required a modification of
    the law of gravity. Briefly, the Modified Newtonian Dynamics (MOND) of
    Milgrom invokes a change in the long-range behavior of gravity: at
    accelerations much smaller than 10$^{-11}$g the force behaves as the
    inverse distance, not as the inverse distance squared. MOND remained a
    theoretical posibility until early this year McGaugh \& de Blok (1998)
    showed that the best fit to the observable properties of LSB disks was
    with MOND, not with Newtonian gravity. Until now, no similar studies
    have been performed specifically for LSB DIGs. It is possible that in
    the MOND regime, which may be best tested in LSB DIGs, the SF may
    proceed in a different manner than in the Newtonian regime. 

    If there is truth in the assumption that there is some kind of link
    between gravity, mass, and light, and that galaxies form in the deeper
    potential wells of the Universe, then the intergalactic space, where gravity is weak,
    should be dominated by a MOND-like behavior of the gravity. Therefore, learning more
    about this theory, and in particular attempting to falsify it, should be
    high on the agenda of present-day astrophysics.

    \section{Conclusions }

    Dwarf galaxies form a natural link between ``normal'' large galaxies and
    the intergalactic matter. They are probably the building blocks out of
    which large galaxies, such as the Milky Way, are formed by numerous
    mergers and bursts of star formation. They form from material in the IG
    space and evolve independently, if not ``harrased'' by interactions
with other objects in
    a dense cluster of galaxies. Their evolution is probably providing the
    IGM with the metals now observed in the X-ray emission of rich clusters,
    but may also provide (part of) the Ly-$\alpha$ blanket of lines seen against
    background QSOs. Finally, the LSB dwarfs may prove observational tests
    of MOND.

        \section*{Acknowledgements}

    I am grateful for continued support of the Center of Excellence in
    Astronomy at Tel Aviv University, the Israel Science Foundation, the
    US-Israel Bi-National Foundation, and the Austrian Friends of Tel Aviv
    University. I acknowledge the various contributions by my former and
    present students to the study of dwarf irregular galaxies, and many discussions
    with friends and colleagues around the world on these subjects.

       %\section*{References}

    \end{document}